\documentclass[aps,prc,twocolumn,superscriptaddress,showpacs,floatfix]{revtex4-1}

\usepackage[T1]{fontenc}
\usepackage{latexsym}
\usepackage{graphicx,amssymb}
\usepackage[fleqn]{amsmath}
\usepackage{amsfonts}
\usepackage{dcolumn}% Align table columns on decimal point
\usepackage{bm}% bold math
\usepackage{braket}
\usepackage{multirow}
\usepackage{MnSymbol}
\usepackage{color}
\usepackage{ulem}
\usepackage{hyperref}
\usepackage{bm}% bold math
\usepackage{color}

\newcommand{\intt}
{
	\displaystyle \int
}

\newcommand{\hatvec}[1]
{
	\hat{ \vec{#1} }
}

\begin{document}

\preprint{ }

\title{Gamow shell model description of the radiative capture reaction $^8$Li$(n,\gamma)$$^9$Li }

\author{G.X. Dong}
\affiliation{School of Science, Huzhou University, Huzhou 313000, China}

\author{X.B. Wang}
\affiliation{School of Science, Huzhou University, Huzhou 313000, China}

\author{N. Michel}
\affiliation{Institute of Modern Physics, Chinese Academy of Sciences, Lanzhou, Gansu 730000, China}

\author{M. P{\l}oszajczak}
\thanks{ploszajczak@ganil.fr}
\affiliation{Grand Acc\'el\'erateur National d'Ions Lourds (GANIL), CEA/DSM - CNRS/IN2P3,
BP 55027, F-14076 Caen Cedex, France}

\date{\today}

\begin{abstract}
\noindent
  {{\bf Background:} The $^8$Li$(n,\gamma)$$^9$Li reaction plays a critical role in several reaction chains leading to the nucleosynthesis of $A>12$ nuclei. Due to unstable nature of $^8$Li and the unavailability of neutron targets, direct measurements of this reaction are exceedingly difficult. Only upper limits of this cross section, provided by the indirect experiments, have been obtained so far. \\
   {\bf Purpose:} In this work, we use the Gamow shell model (GSM) in the coupled-channel representation (GSM-CC) to study the properties of $^9$Li and the radiative capture reaction $^8$Li$(n,\gamma)$$^9$Li.  \\
  {\bf Method:} GSM-CC is a theoretical framework allowing for the description of both nuclear structure and reaction cross sections. In GSM-CC calculations, a translationally invariant Hamiltonian is used with a finite-range two-body interaction tuned to reproduce the low-energy spectra of $^{8-9}$Li.
  The reaction channels are built  by coupling wave functions of the ground state $2_1^+$, the first excited state $1^+_1$, and the first resonance state $3^+_1$ in $^8$Li with the neutron wave function of the projectile in different partial waves. In the calculation of $^8$Li$(n,\gamma)$$^9$Li cross section, all relevant E1, M1, and E2 transitions from the initial continuum states to the final bound states ${3/2}_1^-$, ${1/2}_1^-$ and the resonance ${5/2}_1^-$ of $^9$Li are included.  \\
  {\bf Results:} The GSM-CC approach reproduces the experimental low-energy spectrum, neutron emission threshold, and spectroscopic factors in $^9$Li.
   The calculated reaction rate is consistent with the experimental upper limit of the reaction rate obtained in the indirect measurements at stellar energies.
    \\
  {\bf Conclusion:} The GSM-CC calculations suggest that the $^8$Li$(n,\gamma)$$^9$Li reaction can reduce heavy-element production via the main chain $^7$Li($n,\gamma$)$^8$Li($\alpha,n$)$^{11}$B($n,\gamma$)$^{12}$B($\beta^+$)$^{12}$C. Major contribution to the calculated cross section is given by the direct E1 transition to the ground state of $^8$Li. The contribution of excited states to the reaction rate does not exceed $\sim$18\% of the total reaction rate. }

\end{abstract}

\pacs{25.40.Lw, % Radiative capture
	25.40.Ny, % Resonance reactions
	26.35.+c, % Big Bang nucleosynthesis
	27.20.+n % 6 ¡Ü A ¡Ü 19 (Properties of specific nuclei listed by mass ranges)
    24.10.Cn % Many-body theory
}

\maketitle
\section{Introduction}
\label{intro}

It is well known that the radiative neutron capture reaction $^8$Li$(n,\gamma)$$^9$Li plays an important role in the inhomogeneous big-bang nucleosynthesis and in the r-process. Once $^7$Li is formed, two competing reaction chains paving the way towards production of elements heavier than carbon can occur: $^7$Li($n,\gamma$)$^8$Li($\alpha,n$)$^{11}$B($n,\gamma$)$^{12}$B($\beta^+$)$^{12}$C and $^7$Li($n,\gamma$)$^8$Li($n,\gamma$)$^9$Li($\alpha,n$)$^{12}$B($\beta^+$)$^{12}$C \cite{Malaney88,Boyd92,Rauscher94,xgu95}. The primary reaction chain which begins with $^7$Li($n,\gamma$)$^8$Li($\alpha,n$)$^{11}$B could be significantly reduced if the neutron capture by $^8$Li is large~\cite{Malaney88}.
Thus,  in a neutron-rich environment, the short-lived $^8$Li isotope  may affect the synthesis of heavier elements and the
abundances of $^{9}$Li, $^8$Be, $^{11,12}$B, and $^{12}$C. Once $^{12}$C is reached along the process, heavier nuclei can be produced by $\alpha$ capture \cite{Woosley92,Woosley94}.

Another possibility to produce heavy elements in the $r$ process happens at the beginning of an expansion phase of type II supernovae \cite{Woosley92,Meyer92,Woosley94}. In an environment rich in He isotopes,
the mass gaps at A = 5 and 8 can be overcome by the
$\alpha+\alpha+n \rightarrow ^{9}$Be and $\alpha+\alpha+\alpha \rightarrow ^{12}$C reactions. In addition, if the abundance of neutrons is sufficiently high, the A=8 mass gap can be bridged  by the reaction chain $^4$He($2n,\gamma$)$^6$He($2n,\gamma$)$^8$He($\beta^-$)$^8$Li($n,\gamma$)$^9$Li($\beta^-$)$^9$Be~\cite{Wiescher95,Efros96,Terasawa01}. The competition of this chain of reactions with the decay chain $^8$Li$(\beta^-)$$^8$Be$(2\alpha)$ depends on the neutron abundance and the cross section of $^8$Li$(n,\gamma)$$^9$Li reaction.

Due to the short half life of $^8$Li, the direct measurement of the neutron capture cross section is difficult. Hence, efforts have been made to determine this cross section using indirect measurements.
Upper limits have been obtained in the Coulomb-dissociation method using a $^9$Li beam passing through the virtual photon field of a high-Z nucleus~\cite{Zecher1998,Kobayashi03}.
Several attempts to extract the radiative neutron capture rate for the reaction $^8$Li$(n,\gamma)$$^9$Li have been attempted using transfer reactions to obtain experimental spectroscopic factors (SFs) which were then used to calculate the neutron capture cross section in the potential model~\cite{ZHLi05,Garcia07}. Cross sections obtained in this way agree with the limit given by Zecher et al.~\cite{Zecher1998}, but are significantly higher than the values reported in Ref.~\cite{Kobayashi03}.
The rate of $^8$Li$(n,\gamma)$$^9$Li reaction has also been estimated  in Refs.~\cite{Rauscher94,Malaney89} based on the information existing in other nuclei.

The radiative capture cross sections in mirror reactions: $^8$Li($n,\gamma$)$^9$Li and $^8$B($p,\gamma$)$^9$C have been calculated by Mohr~\cite{Mohr03} using the potential model. It was found that cross sections for these two reactions can be described simultaneously using carefully chosen parameters of the potential. The neutron capture by $^8$Li has also been studied, combining the shell model and the potential model~\cite{Mao91,HLMa12}. However, reaction rates obtained in these studies differ by a factor five.

The microscopic cluster model has been used to investigate the mirror reactions: $^8$Li($n,\gamma$)$^9$Li and $^8$B($p,\gamma$)$^9$C reactions~\cite{Descouvemont93}. The reaction rate calculated in this model is larger than that found in Ref.~\cite{HLMa12}.
Another calculation of the $^8$Li($n,\gamma$)$^9$Li cross section has been done using the modified potential cluster model~\cite{Dubovichenko16}.

Potential model has also been employed to calculate the Coulomb dissociation cross-section of $^9$Li on heavy targets~\cite{Bertulani99,Banerjee08}.
The rate of the neutron capture reaction $^8$Li($n,\gamma$)$^9$Li has been obtained from the Coulomb dissociation cross-section using the principle of detailed balance \cite{Bertulani99,Banerjee08}. The rates obtained in these two studies differ by about 50\%.

Recently, the {\it ab initio} calculation of the cross section $^8$Li($n,\gamma$)$^9$Li  has been done in the no-core shell model with continuum (NCSMC)~\cite{NCSM21}.
The results obtained in this model are significantly higher than most earlier predictions but remain on a higher side of the experimental limit given in Ref.~\cite{Zecher1998}.
In general, most theoretical predictions of considered cross sections are below the upper limit given in Ref.~\cite{Zecher1998}. However, the results obtained with these different approaches might differ by more than one order of magnitude.

In this paper, we will study the low-energy spectra of $^{8,9}$Li and the neutron radiative capture cross-section of the reaction $^8$Li$(n,\gamma)$$^9$Li in the
Gamow shell model framework (GSM)~\cite{Michel02,Michel03,Michel09,Jaganathen14,Fossez15,GSMbook}.
In GSM, many-body states are expanded through a linear combination of Slater determinants spanned by the single particle (s.p.) bound, resonance, and non-resonant scattering states of the Berggren ensemble~\cite{rf:4}. GSM is, in fact, the tool {\it par excellence} for nuclear structure studies. Due to the lack of separation between different channels, GSM cannot be used to describe nuclear reactions. To reconcile this model with reaction theory, GSM had to be formulated in the coupled-channel representation (GSM-CC)~\cite{Jaganathen14,Fossez15,Wang21,GSMbook}.
Matrix elements of the GSM-CC Hamiltonian are calculated in the harmonic oscillator basis, so that the GSM-CC Hamiltonian is Hermitian.
The reaction cross-sections are obtained by coupling the real-energy incoming partial waves with the target states given by the Hermitian Hamiltonian. Consequently, the theoretical framework of the cross section calculation is fully Hermitian, whereas complex energies of resonances arise because the Hamiltonian matrix in Berggren representation is complex symmetric. The GSM-CC approach has been successfully applied to low-energy elastic and inelastic proton scattering reactions~\cite{Jaganathen14}, radiative proton/neutron capture reactions~\cite{Fossez15,dong17}, and deuteron elastic scattering reactions~\cite{Mercenne19}.

The paper is organized as follows. The general formalism of the GSM-CC is briefly introduced in Chap.~\ref{sec-2}. In this chapter, the Hamiltonian is presented in Chap. \ref{sec-2a}, and the GSM-CC coupled-channel equations are discussed shortly in Chap. \ref{sec-2b}.
Results of GSM-CC calculations are discussed in Chap.~\ref{sec3}. In particular, the description of the low-energy spectrum of $^{9}$Li is presented in Chap. \ref{sec3-a}, and the low-energy capture cross-sections of $^8$Li$(n,\gamma)$$^9$Li reaction are analyzed in Chap. \ref{sec3-b}. The obtained reaction rates for different final states of $^9$Li, are presented in Chap. \ref{sec3-c}. Finally, main results are summarized in Chap.~\ref{sec4}.

\section{The Gamow shell model in the coupled-channel representation}
\label{sec-2}

In this chapter, we shall briefly introduce the GSM-CC approach. Details can be found in Ref.~\cite{GSMbook,Michel09} and references cited therein.

\subsection{The Gamow shell model Hamiltonian}
\label{sec-2a}
We will work in the core + valence particles framework in GSM.
The spurious center of mass (CM) excitations in the GSM wave function are removed by rewriting the Hamiltonian in the frame of intrinsic nucleon-core coordinates, i.e.~the cluster-orbital shell model (COSM)~\cite{Ikeda88}:
\begin{equation}
	\hat{H} = \sum_{i = 1}^{ {N}_{ \text{val} } } \left( \frac{ \hat{\vec{p}}_{i}^{2} }{ 2 { \mu }_{i} } + {U}_{\text{core}} ( \hat{r}_{i} ) \right) + \sum_{i < j}^{ {N}_{ \text{val} } } \left( V ( \hat{\vec{r}}_{i} - \hat{\vec{r}}_{j} ) + \frac{ {\hat{\vec{p}}_{i}}{\cdot} {\hat{\vec{p}}_{j} }}{ {M}_{\text{core}} } \right)~,
	\label{GSM_Hamiltonian}
\end{equation}
where $N_\text{val}$ is the number of valence particles, $M_{\text{core}}$ stands for the mass of the core, and $\mu_i$ is the reduced mass of the proton or neutron. $U_{\text{core}}(\hat{r})$ is the s.p.~potential induced by the core acting on each valence nucleon. $V(\hat{\vec{r}}_i-\hat{\vec{r}}_j)$ is the translationally invariant two-body interactions for valence nucleons. The last term in the above formula is the recoil term, which takes into account the finite mass of the core~\cite{GSMbook}.

It is convenient to separate the Hamiltonian into basis and residual parts:
\begin{equation}
{ \hat{H} = \hat{T}+\hat{U}_{ \text{basis} } + (\hat{V}_{ \text{res} } - {\hat{U}_0}) } \; ,
\label{eq_H_GSM}
\end{equation}
In this expression,  ${ \hat{T} }$ is  the kinetic term, ${ \hat{U}_{ \text{basis} } }$ is the potential capturing the bulk properties of the A-particle system and generating the single-particle basis, ${\hat{U}_0} = \hat{U}_{ \text{basis} } - {U}_{\text{core}}$, where
\begin{equation}
	\hat{U}_{ \text{basis} } = \sum_{ i = 1 }^{ {N}_{ \text{val} } } ( {U}_{\text{core}} ( \hat{r}_{i} ) + {\hat{U}_0( \hat{r}_{i} )} )
	\label{eq_U_basis}
\end{equation}
and ${ \hat{V}_{ \text{res} }}$ is the residual interaction
\begin{equation}
	\hat{V}_{ \text{res} } = \sum_{i < j}^{ {N}_{ \text{val} } } \left( V ( \hatvec{r}_{i} - \hatvec{r}_{j} ) + \frac{ {\hatvec{p}_{i}}{\cdot} {\hatvec{p}_{j} }}{ {M}_{\text{core}} } \right) - \sum_{ i = 1 }^{ {N}_{ \text{val} } } {\hat U}_0( \hat{r}_{i} ) \; .
	\label{eq_H_res}
\end{equation}
Note that the $(\hat{V}_{ \text{res} } - {\hat{U}_0})$ operator in (\ref{eq_H_GSM}) is finite-ranged, which is advantageous in numerical calculations.

\subsection{The GSM coupled-channel equations}
\label{sec-2b}
The ${ A }$-body state of the system is decomposed into reaction channels defined as binary clusters:
\begin{equation}
  \ket{ { \Psi }_{ M }^{ J } } = \sum_{ {c} } \int_{ 0 }^{ +\infty } \ket{{ \left( {c} , r \right) }_{ M }^{ J } } \frac{ { u }_{ {c} }^{JM} (r) }{ r } { r }^{ 2 } ~ dr \; .
  \label{scat_A_body_compound}
\end{equation}
The radial amplitude ${ {u}_{ {c} }^{JM}(r) }$ describes the relative motion between target and projectile in a channel $c$.
It is the solution of GSM coupled-channel equations at total angular momentum $ {J} $ and its projection $ {M} $.
The integration variable ${ r }$ in Eq. (\ref{scat_A_body_compound}) is the relative distance between the CM of the target and of the projectile.
The binary-cluster channel states $\ket{ \left( {c} , r \right)^J_M} $ are defined as:
\begin{equation}
  \ket{ \left( {c} , r \right)^J_M}  = \hat{ \mathcal{A}} \ket{ \{ \ket{ \Psi_{ {\rm T} }^{J_{ {\rm T} } }  } }
  \otimes \ket{ {\Psi_{{\rm p}}^{J_{{\rm p}}} \}_{ M }^{J} } } \; .
  \label{channel}
\end{equation}
The channel index $c$ stands for both mass partitions and quantum numbers, where ${\hat{ \mathcal{A}}}$ denotes the antisymmetrization among nucleons pertaining to different clusters.
The states $\ket{\Psi_{ {\rm T} }^{J_{ {\rm T} }} }$ and $\ket{\Psi_{ {\rm p} }^{J_{ {\rm p} }} }$ are the target and projectile states, with their associated total angular momenta ${ { J }_{ {\rm T} } }$ and ${ { J }_{ {\rm p} } }$, respectively.
The used angular momentum coupling reads ${ \mathbf{ J_{\rm A}} = \mathbf{J_{\rm p}} + \mathbf{ J_{{\rm T}} } }$.

The coupled-channel equations can be formally derived from the Schr{\"o}dinger equation: $H \ket{\Psi_{M_{ A }}^{J_{ A }}} = E \ket{\Psi_{M_{ A }}^{J_{ A }}}$, as:
\begin{equation}
	\sumint\limits_{c} \intt_{0}^{ \infty } dr \, {r}^{2} \left( {H}_{ c' , c } ( r' , r ) - E {O}_{ c' , c } ( r' , r ) \right) \frac{ {u}_{c} (r) }{r} = 0 \; ,
	\label{eq_CC_eqs_general}
\end{equation}
where $E$ is the scattering energy of the A-body system, and used kernels read:
\begin{equation}
	{H}_{ c' , c } ( r' , r ) = \braket{ r' , c' | \hat{H} | r , c }
	\label{eq_CC_H_ME_general}
\end{equation}
and
\begin{equation}
	{O}_{ c' , c } ( r' , r ) = \braket{ r' , c' | r , c }
	\label{eq_CC_O_ME_general},
\end{equation}
which are the Hamiltonian and the norm matrix elements in the channel representation, respectively.

The channel state $\ket{ r , c }$ can be constructed using a complete Berggren set of s.p.~states~\cite{rf:4} which includes bound states, resonances, and non-resonant scattering states from the contour in the complex $k$-plane  \cite{Michel02,Michel03,Michel09,GSMbook}:
\begin{equation}
	\ket{ r , c } = \sum_{i} \frac{ {u}_{i} (r) }{r} \ket{ { \phi }_{i}^{ \text{rad} } , c } \; ,
	\label{eq_CC_basis_state_expansion_Berggren_2}
\end{equation}
where ${ \ket{ { \phi }_{i}^{ \text{rad} } , c } = \hat{ \mathcal{A} } ( \ket{ { \phi }_{i}^{ \text{rad} } } \otimes \ket{c} ) }$, ${u}_{i} (r) / r  = { \braket{ { \phi }_{i}^{ \text{rad} } | r }}$, and $\ket{ { \phi }_{i}^{ \text{rad} }}$ is the radial part.
Due to the antisymmetry between projectile and target states, the channel basis states $\ket{ r , c }$ are nonorthogonal, thus leading to a generalized eigenvalue problem.

To solve the generalized eigenvalue problem, we use orthogonal channel basis states ($\ket{ r , c }_{o} = \hat{O}^{ -\frac{1}{2} } \ket{ r , c } \label{eq_CC_non_ortho_to_ortho_channel_states}$) built from the initial non-orthogonal channel basis states:
\begin{equation}
	{}_{\rm o}\braket{ r' , c' | r , c }_{\rm o} = \frac{ \delta ( r' - r ) }{ {r}^{2} } { \delta }_{ c' c }
	\label{eq_CC_ortho_channel_basis_braket}
\end{equation}
where ${ \hat{O} }$ is the overlap operator. The GSM-CC equations~\eqref{eq_CC_eqs_general} are then transformed into:
\begin{align}
	\sumint\limits_{c} \intt_{0}^{ \infty } dr \, {r}^{2} &( {}_{\rm o}\braket{ r' , c' | \hat{H}_{\rm o} | r , c }_{\rm o} - E {}_{\rm o}\braket{ r' , c' | \hat{O} | r , c }_{\rm o} ) \nonumber \\
	&\times {}_{\rm o}\braket{ r , c | { \Psi }_{\rm o} } = 0 \; ,
	\label{eq_CC_eqs_general_clear_ortho}
\end{align}
where
${ \hat{H}_{\rm o} = \hat{O}^{  \frac{1}{2} } \hat{H} \hat{O}^{  \frac{1}{2} } }$, and $ {\ket{ \Psi_{\rm o} } = \hat{O}^{1/2} \ket{ \Psi } }$.

After the substitution ${ \ket{ \Phi } = \hat{O} \ket{ \Psi } }$, the generalized eigenvalue problem of Eq.~\eqref{eq_CC_eqs_general_clear_ortho} becomes a standard matrix eigenproblem:
\begin{equation}
	\sumint\limits_{c} \intt_{0}^{ \infty } dr \, {r}^{2} ( {}_{{\rm o}}\braket{ r' , c' | \hat{H} | r , c }_{\rm o} - E {}_{\rm o}\braket{ r' , c' | r , c }_{\rm o} ) {}_{\rm o}\braket{ r , c | \Phi } = 0 \ .
	\label{eq_CC_eqs_general_clear_ortho_again}
\end{equation}
In the non-orthogonal channel basis, these coupled-channel equations read:
\begin{equation}
	\sumint\limits_{c} \intt_{0}^{ \infty } dr \, {r}^{2} \braket{ r' , c' | \hat{H}_{\rm m} | r , c } \frac{ {w}_{c} (r) }{r} = E \frac{ {w}_{ c' } ( r' ) }{ r' }
	\label{eq_CC_final}
\end{equation}
where ${ \hat{H}_{\rm m} = \hat{O}^{ - \frac{1}{2} } \hat{H} \hat{O}^{ - \frac{1}{2} } }$ and $${w}_{c} (r)/ r \equiv \braket{ r , c | \hat{O}^{ \frac{1}{2} } | \Psi } = {}_{\rm o}\braket{ r , c | \Phi } \ .$$

The coupled-channel equations (\ref{eq_CC_final}) are solved using a numerical method based on the Berggren basis expansion of the Green's function ${ { (H - E) }^{ -1 } }$, that takes advantage of the Berggren basis completeness.
Details of this method can be found in Refs. \cite{Mercenne19,GSMbook}.

\section{Discussion and results}
\label{sec3}
In GSM and GSM-CC calculations, we assume that $^4$He is an inert core. The core potential in the Hamiltonian is mimicked by a Woods-Saxon (WS) central potential, to which a spin-orbit part is added (see Table~\ref{para-1}). For the two-body residual interaction, the Furutani-Horiuchi-Tamagaki (FHT) finite-range force is used~\cite{Furutani78,Furutani79}. The parameters of this interaction (see the Table~\ref{para-2}) are adjusted to reproduce the binding energies of the low-lying states and the neutron separation energies of $^8$Li and $^9$Li.

In GSM-CC calculations, the two-body part of the Hamiltonian from which  the  channel-channel  coupling  potentials are calculated, has been rescaled by multiplicative corrective factors $c(J^{\pi})$ for $J^{\pi}=3/2^-_1, 1/2^-_1, 5/2^-_1, 7/2^-_1$ states to compensate for missing correlations in the GCM-CC wave function, arising due to the omission of non-resonant channels built by coupling the continuum states of $^8$Li with neutron projectile states in different partial waves. These corrective factors are: $c(3/2^-)=1.028, c(1/2^-)=1.06, c(5/2^-)=1.025, c(7/2^-)=1.03$. Note that their effect is minimal as they differ from unity only a by a few percents.
Detailed discussion of the FHT interaction in GSM calculations can be found in Refs.~\cite{Fossez15,dong17}.
\begin{table}[ht!]
\caption{Parameters of the WS potential of the $^4$He core used in the description of $^8$Li and $^9$Li spectra, as well as the neutron-capture cross-section $^{8}\text{Li}(n,\gamma)^{9}\text{Li}$. The radius of Coulomb potential $r_{\text{Coul}}$ is 2.339 fm.\label{para-1}}
\begin{ruledtabular}
\begin{tabular}{ccc}
  Parameter & Protons & Neutrons \\
  \hline
  $a$ & 0.643 fm & 0.631 fm \\
  $R_0$ & 2.062 fm & 2.146 fm \\
  $V_\text{o}(l=0)$ & 47.017 MeV & 44.845 MeV \\
  $V_\text{o}(l=1)$ & 57.141 MeV &33.001 MeV \\
  $V_\text{o}(l=2)$ & -- &33.576 MeV \\
   $V_\text{so}(l=1)$ & 2.814 MeV & 11.296 MeV \\
  $V_\text{so}(l=2)$ & -- & 12.795 MeV \\
\end{tabular}
\end{ruledtabular}
\end{table}

\begin{table}[ht!]
\caption{Parameters of the FHT interaction in GSM and GSM-CC calculations. The superscripts C, SO, and T denote central, spin-orbit, and tensor, respectively. The indices ``s" and ``t" stand for singlet and triplet, respectively \label{para-2}}
\begin{ruledtabular}
\begin{tabular}{cc}
  Parameter & Value \\
    \hline
  $\nu^\textsf{C}_{\textsf{t,t}}$  & -3.190 MeV \\
  $\nu^\textsf{C}_{\textsf{s,t}}$  & -10.956 MeV \\
  $\nu^\textsf{C}_{\textsf{s,s}}$  & -2.932 MeV \\
  $\nu^\textsf{C}_{\textsf{t,s}}$  & -6.858 MeV \\
  $\nu^\textsf{SO}_{\textsf{t,t}}$  & -568.363 MeV \\
  $\nu^\textsf{SO}_{\textsf{s,t}}$  & 0 MeV \\
  $\nu^\textsf{T}_{\textsf{t,t}}$  & -13.538 MeV fm$^{-2}$ \\
  $\nu^\textsf{T}_{\textsf{s,t}}$  & -12.010 MeV fm$^{-2}$ \\
\end{tabular}
\end{ruledtabular}
\end{table}

The Berggren ensemble for neutrons consists of two resonant s.p.~states $0p_{3/2}$ and $0p_{1/2}$, 30 s.p.~states in the non-resonant continuum along the contours $\mathcal{L}^+_{p_{1/2}}$, $\mathcal{L}^+_{p_{3/2}}$, and several bound s.p.~states $s_{1/2}$, $d_{3/2}$ and $d_{5/2}$, mimicking the effects of scattering states of positive energy, so that they are denoted as "scattering-like". Each contour consists of three segments connecting the points: $k_{\text{min}}$=0.0, $k_{\text{peak}}=0.15-i0.14$ fm$^{-1}$, $k_{\text{middle}}$=0.3 fm$^{-1}$ and $k_{\text{max}}$=2.0 fm$^{-1}$, and each segment is discretized with 10 points. Five harmonic oscillator states are considered for each scattering-like partial wave in the $sd$ shell. For protons, only $0p_{3/2}$ and $0p_{1/2}$ resonant s.p.~states are included in the used Berggren basis. Thus, GSM and GSM-CC calculations are performed using 77 basis shells for neutrons: 62 $p_{3/2}$ and $p_{1/2}$ shells, 15 $s_{1/2}$, $d_{3/2}$, and $d_{5/2}$ scattering-like shells, and 2 shells $p_{3/2}$ and $p_{1/2}$ for protons. To reduce the dimension of the Fock space, the basis of Slater determinants is truncated by limiting the occupation of scattering shells and scattering-like shells to two particles.

\subsection{Energy spectrum of $^9$Li}
\label{sec3-a}

For the first step of the GSM-CC calculation, the states of the target nucleus $^8$Li are calculated in GSM. The ground-state energy of $^8$Li with respect to $^4$He equals -13.159 MeV, in good agreement with the experimental value -12.982 MeV~\cite{Tilley04}.
The calculated energy of the first resonance is -11.004 MeV and its width equals 70.1 keV, close to the experimental resonance energy -10.727 MeV and width 33(6) keV)~\cite{Tilley04}.

\begin{figure}[htb]
\includegraphics[width=0.9\linewidth]{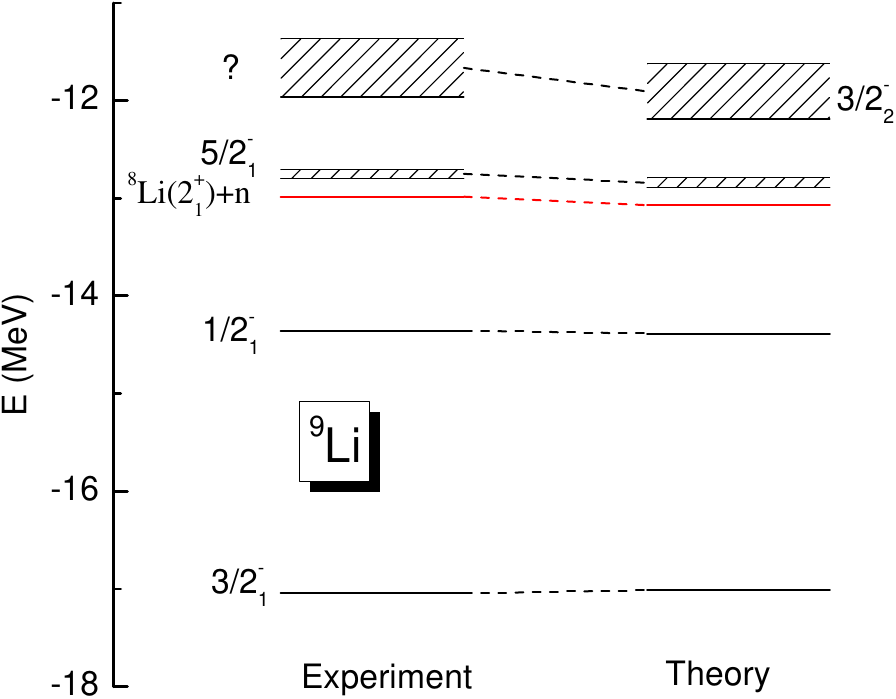}
\caption{GSM-CC energy levels of $^9$Li are compared with experimental data~\cite{Tilley04}. Energies are given relatively to the binding energy of $^4$He.}
\label{fig-1b}
\end{figure}

The channel states in GSM-CC are built by coupling the ground state $J_{\text{T}}^\pi=2^+_1$, the first excited state $J_{\text{T}}^\pi=1^+_1$ and the lowest-energy resonance state $J_{\text{T}}^\pi=3^+_1$ of $^8$Li with the neutron projectile in the following partial waves: $s_{1/2}$, $p_{1/2}$, $p_{3/2}$, $d_{3/2}$ and $d_{5/2}$. The GSM-CC energies and widths of the states $[^8\text{Li}(J_{\text{T}}^\pi)\otimes \nu_{\ell_j}]^{J_f^\pi}$of $^9$Li, and the neutron separation energy are compared with experimental data~\cite{Tilley04} in Fig.~\ref{fig-1b}.
\begin{table}[ht!]
\caption{GSM-CC excitation energies of low energy bound and resonance states in $^{9}$Li are compared with experimental data. All energies are given relatively to their respective ground states. Experimental data are taken from Ref.~\cite{Tilley04}.\label{spectra}}
\begin{ruledtabular}
\begin{tabular}{c|c|c|c|c|c}
 \multicolumn{3}{c|}{Theory}& \multicolumn{3}{c}{Experiment} \\
   \hline
   J$^{\pi}$ & E(MeV) & $\Gamma$(keV) &  J$^{\pi}$ &   E(MeV) & $\Gamma$(keV)  \\
    \hline
  ${3/2}^-_1$  &0.000 & - &  ${3/2}^-_1$& 0.000& -\\
        ${1/2}^-_1$  &2.625 & - &  ${1/2}^-_1$& 2.691& -\\
        ${5/2}^-_1$  &4.176 & 105.7 &  ${5/2}^-_1$& 4.296& 100(30)\\
        ${3/2}^-_2$  &5.109 & 568.1 &   ? & 5.38 & 600(100)\\
        ${3/2}^-_3$  &5.219 & 911.0 & ? &6.43 & 40(20)\\
         ${7/2}^-_1$  &6.433 &173.5 & & &\\
\end{tabular}
\end{ruledtabular}
\end{table}

One can see  that calculated energy levels and neutron separation energies are in good agreement with experimental data, especially for the low-lying bound states ${3/2}^-_1$ and ${1/2}^-_1$. The resonance with unknown spin and parity in $^9$Li is predicted to be a $J^{\pi}=3/2^+_2$ state. Excitation energies and widths of low-lying states in $^{9}$Li are listed in Table~\ref{spectra}.

As seen in Fig.~\ref{fig-1b}, the final nucleus ${ {}^{9}\text{Li} }$ has two states below the one-neutron emission threshold: ${3/2}_{1}^{-}$ and ${1/2}_{1}^{-}$. The calculated neutron separation energy in ${ {}^{9}\text{Li} }$ is ${ {S}_{ \text{n} }^{ ( \text{th} ) } = 3.945}$ MeV, which agrees well with the experimental value ${ {S}_{ \text{n} }^{ ( \text{exp} ) } = 4.062}$ MeV.

Even though it is not a physical observable \cite{Furnstahl02,Duguet2015,Gomez2021}, SFs are useful as they capture information on configuration mixing in the many-body wave function. The SFs $\langle ^9{\rm Li}(J^{\pi})|[^8{\rm Li}_{\rm g.s.}(2^+_1)\otimes \nu_{\ell_j}]\rangle$ calculated in GSM for ${3/2}^-_1$, ${1/2}^-_1$, ${5/2}^-_1$ states in $^9$Li are compared in Table~\ref{sp-factor} with results issued from the variational Monte Carlo approach \cite{Wuosmaa05}, the no-core shell model (NCSM) \cite{Navratil04},
the shell model (SM) \cite{Fortune19}, and experimental data~\cite{Wuosmaa05}.

 In GSM, SFs are defined as:
\begin{equation} {\label{eq:spectroscopic_factor}}
{\cal S}^2_{\ell_j} = \int\hspace{-1.4em}\sum {{\cal A}^2_{\ell_j}(k_n)} \; ,
\end{equation}
where
\begin{equation} {\label{eq:spectroscopic_amplitude}}
{\cal A}_{\ell_j}(k_n) = \langle \Psi_{A} || a^+_{\ell_j}(k_n) || \Psi_{A-1}\rangle/\sqrt{2 J_A + 1}
\end{equation}
are the spectroscopic amplitudes~\cite{Michel2007,Michel2007a}, $\Psi_{A}$ is the wave function of the A-nucleon system, $J_A$ is its total angular momentum, and $a^+_{\ell_j}(k_n)$ is a nucleon  creation operator associated with the Berggren basis state $|k_n\rangle$.
Eq.\,(\ref{eq:spectroscopic_factor}) involves a summation over discrete resonant states and scattering-like states, and an integration along the contour of scattering states in the Berggren ensemble. The spectroscopic factor is  defined as the sum of squared spectroscopic amplitudes associated with possible reaction channels. For instance, for the neutron SF:
$[^{9}{\rm Li}_{\rm g.s.}({3/2}^-_1)\rightarrow {^{8}{\rm Li}_{\rm g.s.}(2^+_1)} + {\rm n}]$, both $p_{3/2}$ and $p_{1/2}$ partial waves contribute and their SFs are added.
\begin{table}[ht!]
\caption{Neutron spectroscopic factors $\langle ^9{\rm Li}(J^{\pi})|[^8{\rm Li}_{\rm g.s.}(2^+_1)\otimes \nu_{\ell_j}]\rangle$
in the low-lying states of $^9$Li. See text for details.}
\label{sp-factor}
\begin{ruledtabular}
\begin{tabular}{c|cccc|c}
  J$^{\pi}$ & VMC\cite{Wuosmaa05}& NCSM\cite{Navratil04}& SM\cite{Fortune19} & GSM& $(d,p)$\cite{Wuosmaa05}\\
    \hline
  ${3/2}^-_1$  & 1.11 & 1.05 & 0.93& 0.93& 0.90(13)\\
  ${1/2}^-_1$  & 0.52 & 0.52 & 0.38& 0.42&0.73(15)\\
  ${5/2}^-_1$  & 0.78 & 0.84 & 0.79& 0.80 &0.93(20)\\

\end{tabular}
\end{ruledtabular}
\end{table}

There is clearly an agreement between the different theoretical approaches presented.
For the ${3/2}^-_1$ and ${5/2}^-_1$ states of $^9$Li, all models reproduce the experimental value of SFs. However, for the first excited state ${1/2}^-_1$, all models predict a smaller value than that reported experimentally.

\subsection{Cross section for $^{8}\text{Li}(n,\gamma)^{9}\text{Li}$ reaction}
\label{sec3-b}

Once the initial GSM wave function of $^8$Li is calculated, the neutron capture cross section to the final state of $^9$Li with the total angular momentum ${ {J}_{f} }$ can be calculated from:
\begin{equation}
	{ \sigma }_{ {J}_{f} } ( {E}_{ \text{c.m.} } ) = \intt_{0}^{ 2 \pi } d{ \varphi }_{ \gamma } \intt_{0}^{ \pi } \sin{ \theta }_{ \gamma } d{ \theta }_{ \gamma } \frac{ d{ \sigma }_{ {J}_{f} } ( {E}_{ \text{c.m.} } , { \theta }_{ \gamma } , { \varphi }_{ \gamma } ) }{ d{ \Omega }{ \gamma } } \; ,
	\label{eq_rad_cap_partial_cross_section}
\end{equation}
where $E_{\text{c.m.}}$ is the center of mass energy of $^9$Li, and $E_{\text{c.m.}} = 0$ corresponds to the neutron emission threshold in $^9$Li. The total cross section is then:
\begin{equation}
	\sigma ( {E}_{ \text{c.m.} } ) = \sum_{ {J}_{f} } { \sigma }_{ {J}_{f} } ( {E}_{ \text{c.m.} } ) \; .
	\label{eq_rad_cap_total_cross_section}
\end{equation}
The differential cross-sections ${ d{ \sigma }_{ {J}_{f} } }/{ d{ \Omega }{ \gamma } }$  in (\ref{eq_rad_cap_partial_cross_section}) are calculated using the matrix elements of the electromagnetic operators between the antisymmetrized initial and final states of $^8$Li and $^9$Li, respectively~\cite{Fossez15,dong17,GSMbook}.
The electromagnetic transitions connect the continuum states of $^9$Li with the final states, which are either bound ($J_f = {3/2}_1^-, {1/2}_1^-$) or resonance ($J_f = {5/2}_1^-$).
For E1 transitions, we consider many-body composite continuum states coupled to $J_i = {1/2}^+, {3/2}^+, {5/2}^+, {7/2}^+$.
For M1 and E2 transitions, the composite continuum states entering the reaction are coupled to $J_i = {{1/2}^-, 3/2}^-,  {5/2}^-, {7/2}^-$.

In the calculation of E1 and E2 electromagnetic transitions, effective charges are used. For the E1 transitions, the empirical value of the effective charge is~\cite{Hornyak75,YKHo88}:
\begin{equation}
	{ {e}_{ \text{eff} }^{p}({\rm E1}) = e f_{\rm E1} \left( 1 - \frac{Z}{A} \right) \ ; }\qquad
    {e}_{ \text{eff} }^{n}({\rm E1}) = - e f_{\rm E1} \frac{Z}{A} \; ,
	\label{eq1}
\end{equation}
where ${ Z }$ and ${ A }$ are the proton number and the total number of nucleons in the system formed by the target nucleus and the projectile. The theoretical value for low-energy neutron radiative capture reactions is obtained by imposing $f_{\rm E1} = 1$~\cite{Hornyak75,YKHo88}, which is the effective charge from the recoil effect of the center of mass.
In the study of the low-energy neutron radiative capture reactions, the E1 strength is to a large extent decoupled from the giant dipole state, resulting in a very small core polarization correction~\cite{Lane1961,Allen1979}. So, for this reaction, only recoil effects of the center of mass remain. However, certain studies of the neutron capture reactions in the valence capture model suggest a broader range of neutron effective charges with: $1 \leq f_{\rm E1} < 3$~\cite{YKHo88}. In the following, if not mentioned otherwise, we will use the theoretical value of the effective neutron charge with $f_{\rm E1} = 1$.

 %However, certain studies of the neutron capture reactions in the valence capture model suggest a broader range of neutron effective charges with: $1 \leq f_{\rm E1} < 3$~\cite{YKHo88}. In the following, if not mentioned otherwise, we will use the theoretical value of the effective neutron charge with $f_{\rm E1} = 1$.

For E2 transitions, the proton effective charge is usually in the range of $1.1 e < {e}_{ \text{eff} }^{p}({\rm E2}) < 1.5 e$, and the neutron effective charge is in the range of $0.5 e < {e}_{ \text{eff} }^{n}({\rm E2}) < 0.6 e$ ~\cite{Prestwich84,Castel86,Hees1988,Rydt2009}. However, as will be shown below, the large uncertainty in the values of ${e}_{ \text{eff} }^{p}({\rm E2})$, ${e}_{ \text{eff} }^{n}({\rm E2})$ does not impact significantly the prediction of $^{8}\text{Li}(n,\gamma)^{9}\text{Li}$ reaction rate at astrophysical energies. For M1 transitions, no effective charges are needed.

In Table~\ref{q-moment} we show GSM results for the electric quadrupole and magnetic moments of the ground state of $^{8,~9}$Li. Electric quadrupole moments are calculated for different sets of effective charges. The effective charges $(e_{ \text{eff} }^p, e_{ \text{eff} }^n) = (1.26$e$,~0.47$e$)$ have been determined by a simultaneous optimization of energy levels and static moments in 0$p$-shell nuclei using the standard shell model~\cite{Hees1988}. $(e_{ \text{eff} }^p, e_{ \text{eff} }^n) = (1.1$e$,~0.5$e$)$ have been obtained by a comparison of experimental quadrupole moments of odd-$Z$ even-$N$ isotopes in $(sd)$-shell nuclei with shell model calculation~\cite{Rydt2009}. The effective charges of $(e_{ \text{eff} }^p, e_{ \text{eff} }^n) = (1.5$e$,~0.5$e$)$ have been adopted in the shell model calculation of $fp$-shell nuclei~\cite{Honma04}. For a comparison, we show also GSM results obtained using the bare charges $(e^p, e^n) = (1, 0)$, and the effective charges from the recoil effect of the center of mass $(e_{ \text{eff} }^p, e_{ \text{eff} }^n) =\left( e \left( 1 - \frac{2}{A} + \frac{Z}{ {A}^{2} } \right), eZ/A^2\right)$~\cite{Hornyak75,YKHo88}, labeled as ``recoil correction''). The best agreement with the data in GSM is found using the bare charges. This tendency is expected due to large model space in the GSM calculation.

\begin{table}[ht!]
\caption{The electric quadrupole and magnetic moments of the ground state of $^{8,~9}$Li. The experimental data are from Refs.~\cite{Tilley04,Stone16}. In the GSM calculation of electric quadrupole moment, the results are given for different effective charges $(e_{ \text{eff} }^p, e_{ \text{eff} }^n)$ for protons and neutrons. For more details, see a description in the text.  }
\label{q-moment}
\begin{ruledtabular}
\begin{tabular}{lcc}
  & $^{8}$Li,~${2}^+_1$ & $^{9}$Li,~${3/2}^-_1$\\
    \hline
  $Q\left(e^2 \textrm{fm}^2 \right)$  &  &  \\
  Experiment  & +3.14(2) & -3.04(2) \\
  GSM~(recoil correction) & +2.63 &-2.91 \\
  GSM~(1$e$,~0$e$)  & +3.06  & -3.44  \\
  GSM~(1.26$e$,~0.47$e$)~\cite{Hees1988}  & +5.75  &-5.64   \\
  GSM~(1.1$e$,~0.5$e$)~\cite{Rydt2009}  & +5.38  &-5.17   \\
  GSM~(1.5$e$,~0.5$e$)~\cite{Honma04}  & +6.60  &-6.55   \\
  \hline
 $\mu \left(\mu_\textsc{N} \right)$  &  &  \\
  Experiment  & 1.654 & 3.437\\
  GSM  & 1.618  & 3.014  \\
\end{tabular}
\end{ruledtabular}
\end{table}

Detailed descriptions of the cross-section calculation and various approximations in GSM-CC approach, such as the long-wavelength approximation and the treatment of antisymmetry in the many-body matrix elements of the electromagnetic operators, can be found in Refs.~\cite{Fossez15,dong17,GSMbook}.

The experimental information about $^{8}\text{Li}(n,\gamma)^{9}\text{Li}$ radiative neutron capture cross section is indirect and comes either from the Coulomb dissociation of $^9$Li on heavy targets \cite{Zecher1998,Kobayashi03}, or from transfer reactions \cite{Garcia07,ZHLi05}.
In the measurement provided by the Coulomb dissociation method~\cite{Zecher1998}, the upper limits of direct neutron capture cross section were obtained in two decay energy bins: $E_n \in [0.0, 0.5]$ MeV and [0.5, 1.0] MeV. An upper limit of the direct reaction rate in this experiment is one order of magnitude larger than that reported in another Coulomb-dissociation experiment by Kobayashi et al~\cite{Kobayashi03}. It is also almost 50\% larger than the value extracted from transfer reactions using the potential model and experimental SFs~\cite{Garcia07,ZHLi05}.

\begin{table}[ht!]
\caption{The GSM-CC direct neutron radiative capture cross section to the ground state $J^{\pi}=3/2^-$ of $^9$Li is compared to the experimental data~\cite{Zecher1998}. % for the theoretical value ($f_{\rm E1} = 1$) of the E1 effective charges (\ref{eq1}).
Contributions to the cross section from the neutron capture to the first excited bound state $1/2^-_1$ and the first resonance $5/2^-_1$  are shown as well. Experimental and theoretical cross sections are averaged in the two decay energy bins: $E_n \in [0.0,0.5]$ MeV and $E_n \in [0.5,1.0]$ MeV, and assigned to the representative energies ${\tilde E}_n = 0.25$ and 0.75 MeV, respectively.
\label{sigmabar}}
\begin{ruledtabular}
\begin{tabular}{c|r|r|c|c|c}
  \multicolumn{3}{c|}{ GSM-CC}&\multicolumn{3}{c}{Experiment~\cite{Zecher1998}} \\
   \hline
\multicolumn{1}{c}{${\tilde E}_n$} &\multicolumn{1}{c}{ 0.25 MeV }& 0.75 MeV & \multicolumn{1}{c}{${\tilde E}_n$}  & \multicolumn{1}{c}{0.25 MeV} & 0.75 MeV \\
\hline
 {$J^{\pi}$ } & $\sigma$ ($\mu$b) & $\sigma$ ($\mu$b)& {Target} &  $\sigma$ ($\mu$b) & $\sigma$ ($\mu$b)  \\
    \hline
  $3/2^-$  & 15.23 & 5.24 & U & 19.1$\pm$10.4& 7.9$\pm$5.5 \\
    &    &      & Pb& 17.8$\pm$11.9& 8.1$\pm$6.4 \\
\hline
   $1/2^-$   & 0.80  & 0.59 &  &   & \\
   $5/2^-$   & 0.54 & 1.11  &  &   & \\
\end{tabular}
\end{ruledtabular}
\end{table}
Table \ref{sigmabar} compares the experimental upper limit of the direct neutron radiative capture cross section to the ground state $3/2^-_1$ of $^9$Li~\cite{Zecher1998} with the cross section calculated in GSM-CC.
The contribution of excited states $J^{\pi}={1/2}_1^-$, ${5/2}_1^-$ to the neutron radiative capture cross section is shown separately. The GSM-CC cross sections are obtained for standard values of the E1 and E2 neutron effective charges.
All relevant E1, M1, E2 transitions to the final states $J_f = {3/2}_1^-$, $1/2^-_1$ and $5/2^-_1$ are included. One can see that the contribution of excited states is not negligible, and vary from $\sim$ 9\% of the total neutron capture cross section in the first energy bin $E_n \in [0.0,0.5]$ MeV to 24\% in the second bin $E_n \in [0.5,1.0]$ MeV. For scaling the E1 effective charges $f_{\rm E1}$ of 1.8 or larger values, the calculated direct capture cross section can be larger than the experimental upper limit.

\begin{figure}[htbp]
	\includegraphics[width=0.9\linewidth]{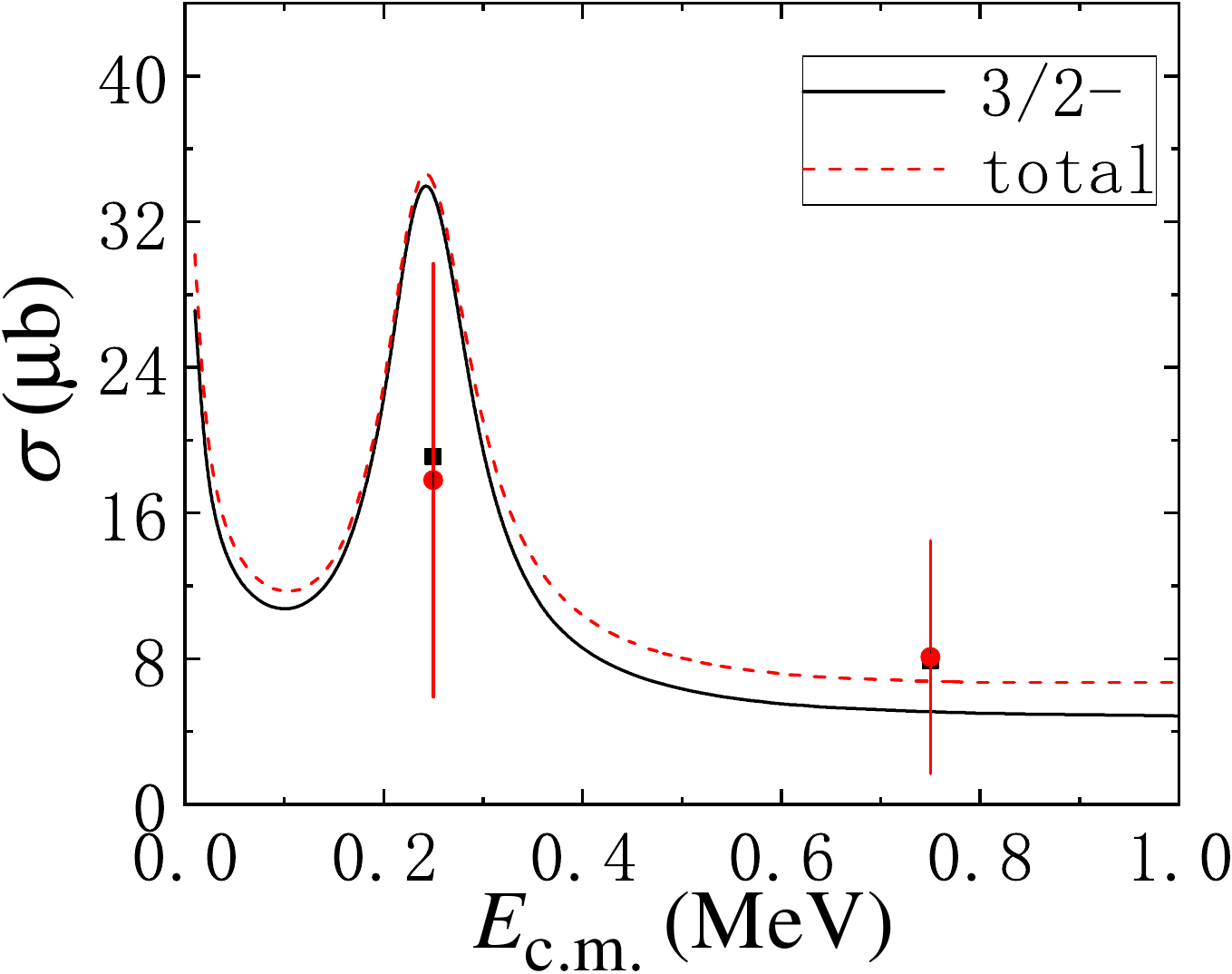}
	\caption{ The GSM-CC neutron radiative capture cross section of the reaction ${ {}^{8}\text{Li} ( n , \gamma ) {}^{9}\text{Li} }$ is plotted as a function of the neutron projectile energy in the n + $^8$Li center of mass frame. The solid line shows the direct capture to the ground state $J^{\pi}={3/2}_1^-$ of $^9$Li and the red dashed line exhibits the total neutron radiative capture cross section which is a sum of contributions from the capture to $J^{\pi}=3/2^-_1, 1/2^-_1$ and $5/2^-_1$ final states.
	All lines represent the fully antisymmetrized GSM-CC results in the longwavelength approximation~\cite{Fossez15,dong17,GSMbook}.
The red points and black squares are the upper limits obtained in the Coulomb-dissociation experiment with Pb and U targets,  respectively~\cite{Zecher1998}. Experimental cross sections at ${\tilde E}_n = 0.25$ MeV and 0.75 MeV correspond to average cross sections in the two decay energy bins: $E_n \in [0.0,0.5]$ MeV and $E_n \in [0.5,1.0]$ MeV.
}
	\label{fig-5}
\end{figure}

Fig. \ref{fig-5} compares direct and total neutron radiative capture cross sections calculated in GSM-CC. % The calculation is done for the theoretical value of the effective charges ($f_{\rm E1} = 1$)~(\ref{eq1}).
In the total neutron capture cross section, all relevant E1, M1 and E2 transitions in the capture to the  $J^{\pi}=3/2^-_1, 1/2^-_1,5/2^-_1$ final states are added up. The experimental upper limits~\cite{Zecher1998} are also listed in the figure.
It is seen that the GSM-CC calculation is consistent with the upper limit given in Ref.~\cite{Zecher1998}.
\begin{figure}[htbp]
	\includegraphics[width=0.9\linewidth]{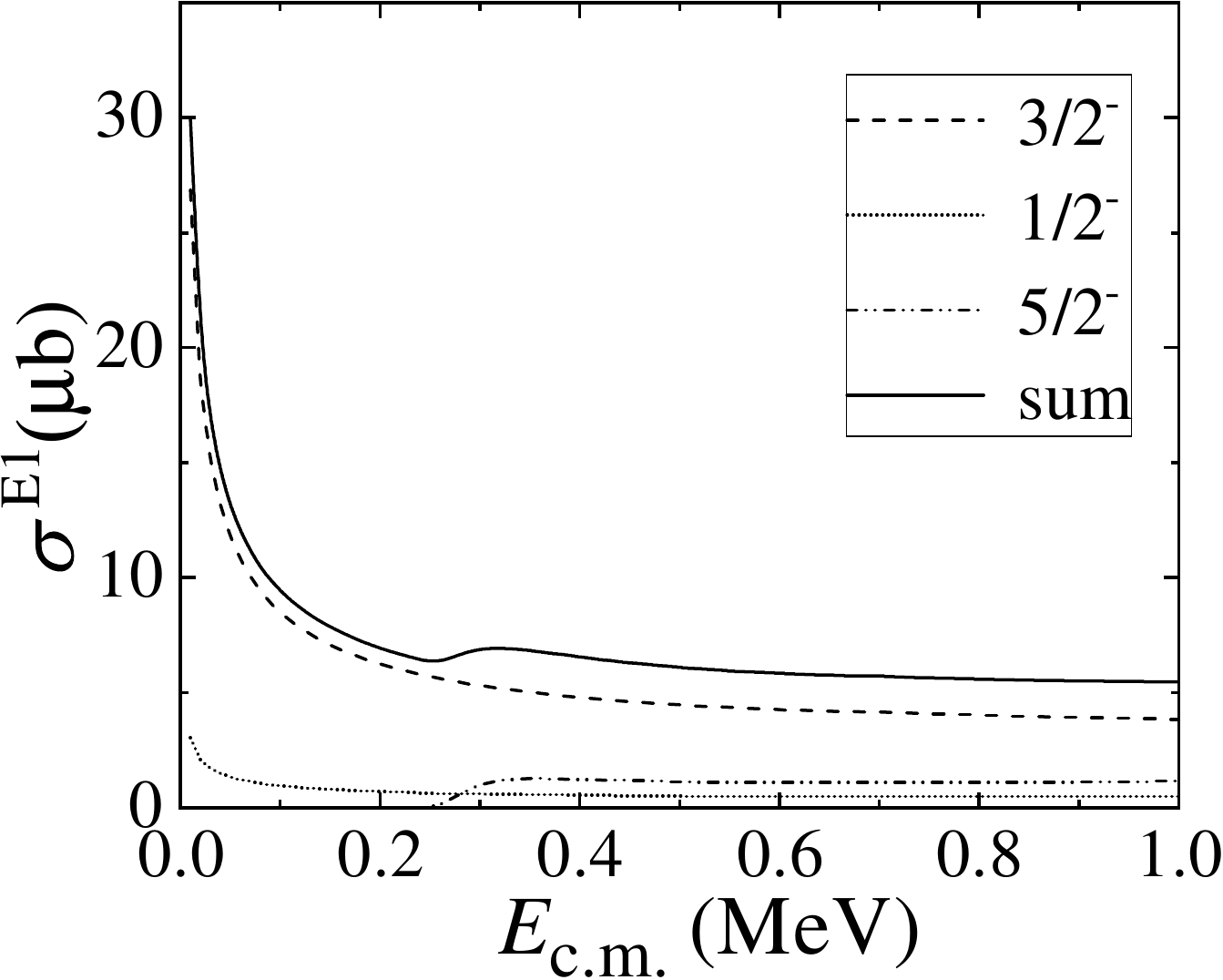}
	\caption{The E1 neutron capture cross section for the $^8\text{Li}(n,\gamma)^9\text{Li}$ reaction is plotted as a function of the neutron projectile energy in the n + $^8$Li center of mass frame. The solid line represents the fully antisymmetrized GSM-CC calculation for the radiative neutron capture to both ground state $J^{\pi}={3/2}_1^-$ and excited states $J^{\pi}={1/2}_1^-, {5/2}_1^-$ of $^9$Li. The dashed, dotted, and the dashed-dotted-dotted lines show the separate contributions to the E1 capture cross section of various final states. }
	\label{fig-2}
\end{figure}
\begin{figure}[htbp]
	\includegraphics[width=0.9\linewidth]{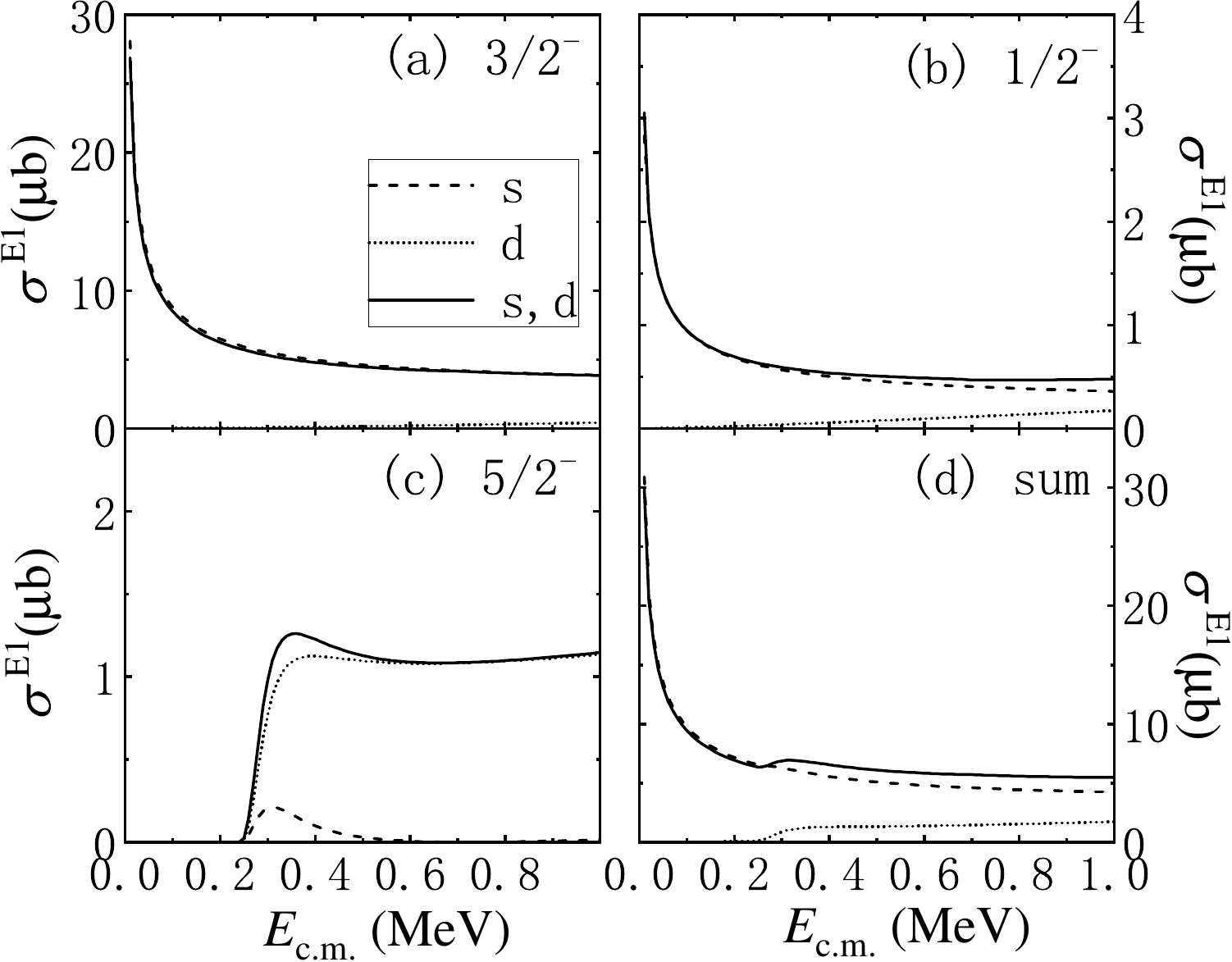}
	\caption{The E1 neutron capture cross section to different final states $J^{\pi}={3/2}_1^-$ (panel (a)), ${1/2}_1^-$ (panel (b), and ${5/2}_1^-$ (panel (c)) in the reaction $^8\text{Li}(n,\gamma)^9\text{Li}$. Panel (d) presents the sum of the E1 capture cross section in panels (a) - (c). For each final state separately, the E1 neutron capture cross section is plotted for $s$-wave neutrons (dashed line), $d$-wave neutrons (dotted line), and a sum of $s$- and $d$-wave contributions (solid line). For more information, see the caption of Fig. \ref{fig-2}.
	 }
	\label{fig-2b}
\end{figure}
\begin{figure}[htbp]	
	\includegraphics[width=0.9\linewidth]{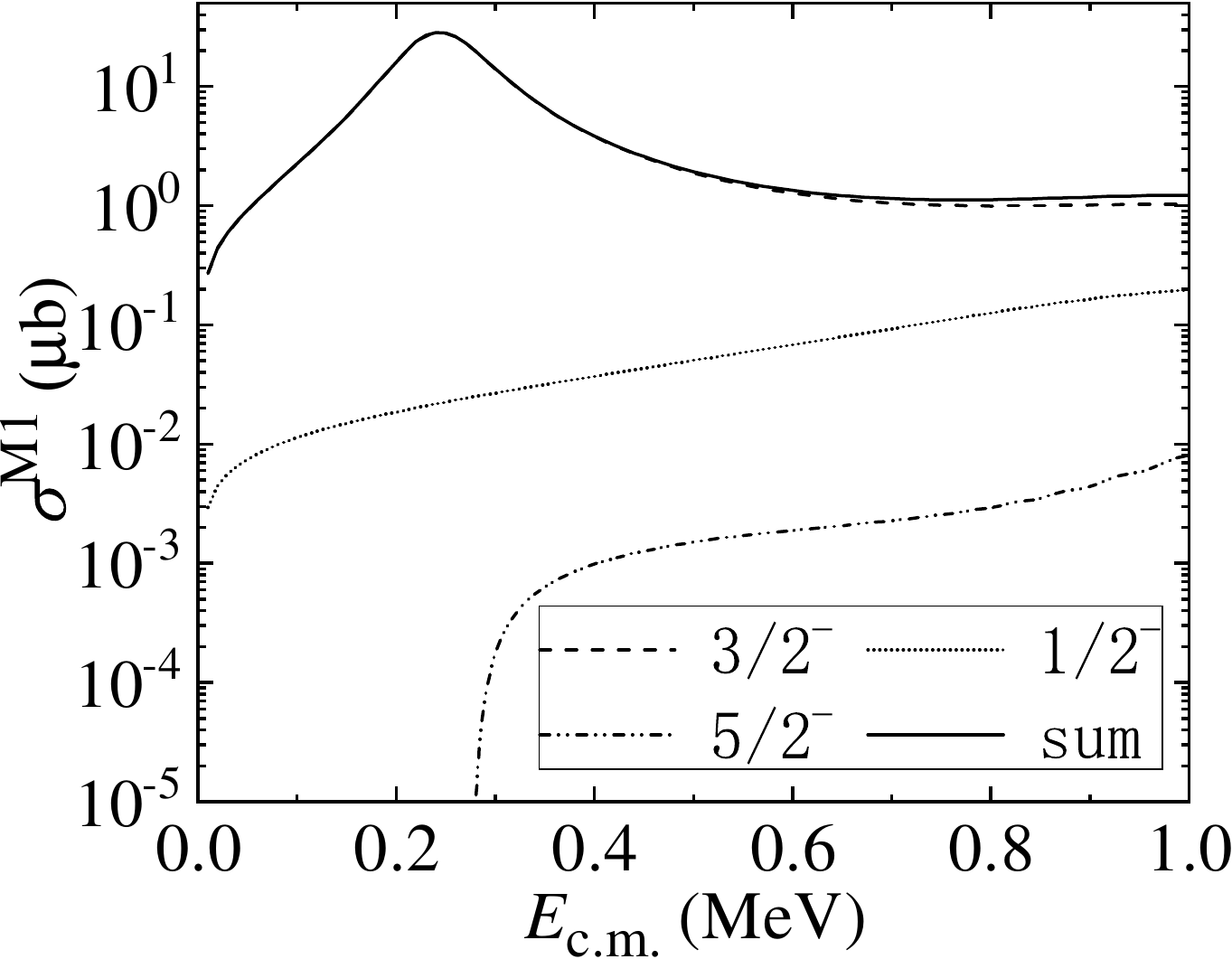}
	\caption{The M1 neutron capture cross section for the $^8\text{Li}(n,\gamma)^9\text{Li}$ reaction.
	The peak corresponds to the ${5/2}_{1}^-$ resonance of $^9\text{Li}$. For more information, see the caption of Fig. \ref{fig-2}. }
	\label{fig-3}
\end{figure}
\begin{figure}[htbp]
	\includegraphics[width=0.90\linewidth]{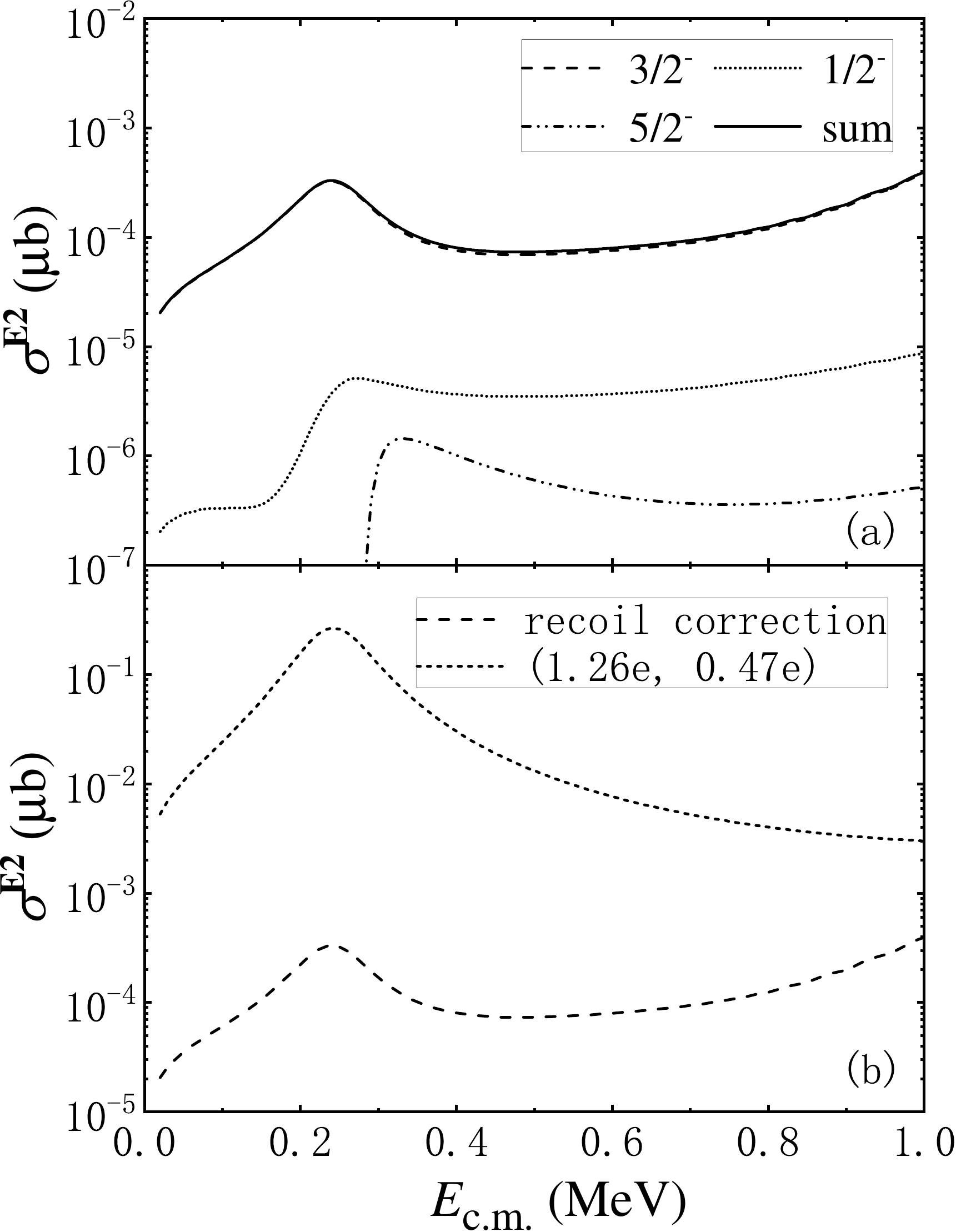}
	\caption{The E2 neutron capture cross section for the $^8\text{Li}(n,\gamma)^9\text{Li}$ reaction. The upper panel (a) shows the separate contributions to the different final states of $^9$Li and their sum.
  The lower panel (b) shows the the sum of contributions to $J^{\pi}=3/2^-_1, 1/2^-_1, 5/2^-_1$ states calculated for two sets of E2 effective charges: $(e_{ \text{eff} }^p, e_{ \text{eff} }^n) = (1.26$e$,~0.47$e$)$~\cite{Hees1988}, and $(e_{ \text{eff} }^p, e_{ \text{eff} }^n) =\left( e \left( 1 - \frac{2}{A} + \frac{Z}{ {A}^{2} } \right), eZ/A^2\right)$~\cite{Hornyak75,YKHo88} resulting from the recoil correction. The peak corresponds to the ${5/2}_{1}^-$ resonance.}
		\label{fig-4}
\end{figure}

Figs.~\ref{fig-2}-\ref{fig-4} show separate contributions to the total cross section in $^8\text{Li}(n,\gamma)^9\text{Li}$ reaction: $\sigma^{\rm E1}$ for E1 transitions (Fig.~\ref{fig-2} and \ref{fig-2b}), $\sigma^{\rm M1}$ for M1 transitions (Fig.~\ref{fig-3}), and $\sigma^{\rm E2}$ for E2 transitions (Fig.~\ref{fig-4}). The largest contribution to the radiative neutron capture cross section comes from a direct capture of $s$-wave neutrons to the ground state $J^{\pi}=3/2^-_1$. This finding agrees with the results issued from other studies~\cite{Dubovichenko16,NCSM21}.

E1 transitions provide the largest part of the cross section besides from the $5/2^-_1$ resonance, where M1 transitions are dominant.
The capture to the first excited state $1/2^-_1$ increases the value of $\sigma^{\rm E1}$ by $\sim 11\%$ (see Fig. \ref{fig-2}). One may notice that the contribution of $5/2^-_1$ resonance is large at energies above $E_{\rm c.m.} \sim 0.35$ MeV. Moreover, the capture of $d$-wave neutrons becomes important at higher energies ($E_{\rm c.m.} > 0.35$ MeV). As seen in Fig. \ref{fig-4}, the contribution of E2 transitions to the neutron capture cross section is minor both for the standard values of E2 effective charges for bound states $(e_{ \text{eff} }^p, e_{ \text{eff} }^n) = (1.26$e$,~0.47$e$)$~\cite{Hees1988}, which can be considered as an upper limit, and for the effective charges from the recoil of the center of mass~\cite{Hornyak75,YKHo88}.

As can be seen in Figs.~\ref{fig-3}-\ref{fig-4}, the observed peak is caused by the resonance ${5/2}_1^-$.
The ${ {5/2}_{1}^{-} }$ resonance, which lies above the one-neutron decay threshold, should be seen in M1 and E2 transitions.
The CM energy of the peak is given by ${E}_{ \text{c.m.} } = {E}_{i}^{ ( A ) } [ \text{GSM-CC} ] - {E}_{0}^{ ( A - 1 ) } [ \text{GSM} ]
\label{eq_rad_cap_resonance_position}$.
${ {E}_{i}^{ ( A ) } [ \text{GSM-CC} ] }$ stands for the GSM-CC energy of resonance ${ i }$ in $^9$Li, and ${ {E}_{0}^{ (A) } [ \text{GSM} ] }$ is the GSM ground-state energy of $^8$Li.
Although the $\sigma^{\rm M1}$ is negligible at very low energies, its contribution can not be neglected in the region of the ${5/2}_1^-$ resonance.
Around the resonance peak, $\sigma^{\rm M1}$ becomes comparable to $\sigma^{\rm E1}$ or even larger. $\sigma^{\rm E2}$ is significantly smaller than both $\sigma^{\rm E1}$ and $\sigma^{\rm M1}$ by several orders of magnitude.
For M1 and E2 transitions, the cross section is generated mostly from the capture to the ground state of ${^9}\text{Li}$.

\subsection{The astrophysical reaction rate}
\label{sec3-c}

\begin{figure}[htbp]
	\includegraphics[width=0.9\linewidth]{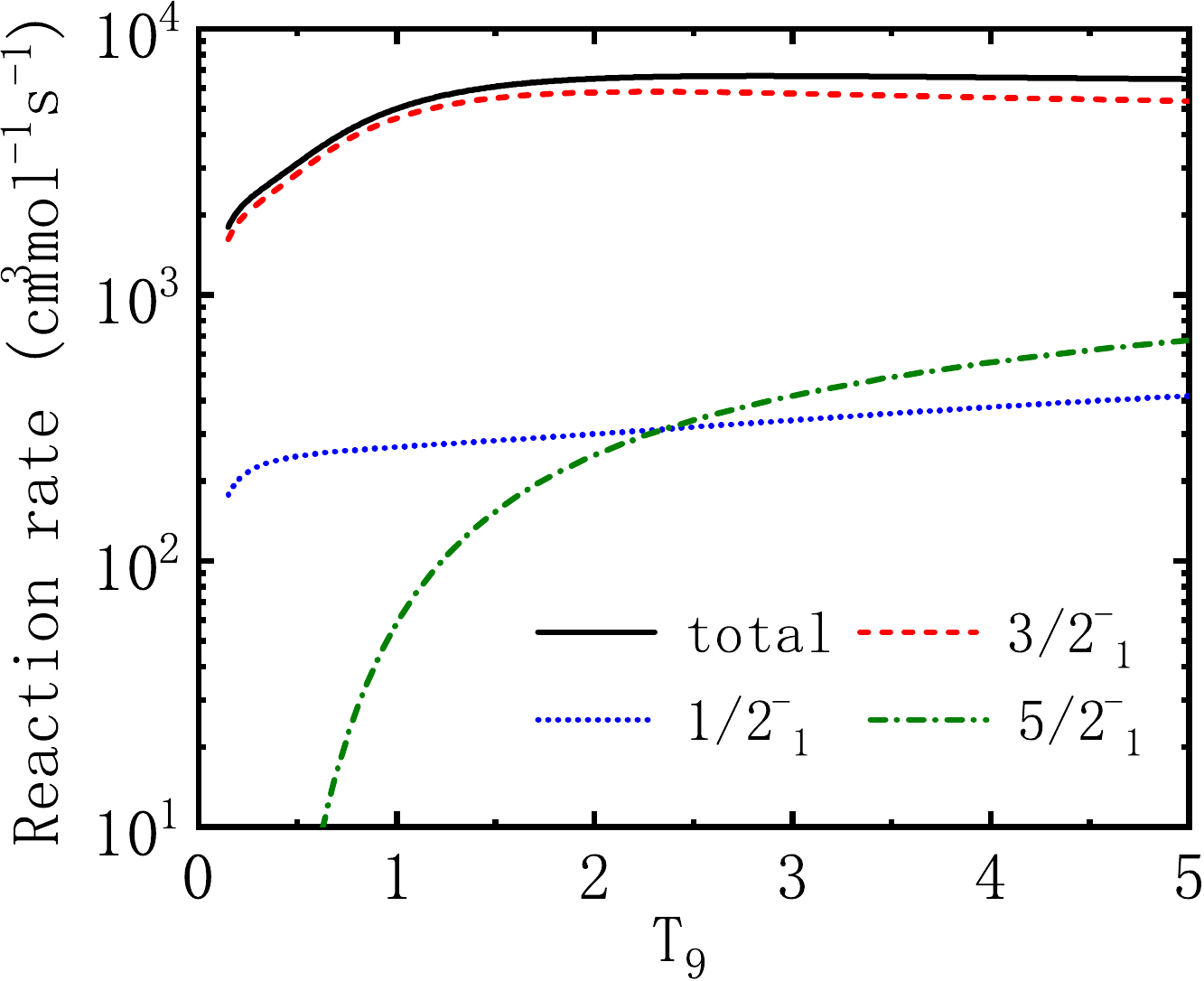}
	\caption{The rate of the ${ {}^{8}\text{Li} ( n , \gamma ) {}^{9}\text{Li} }$ reaction calculated in GSM-CC is shown as a function of temperature ${\rm T}_9$. The total reaction rate is depicted by the solid line. The separate contributions from the ground state $3/2^-_1$ and excited states $1/2^-_1$, $5/2^-_1$ are shown by dashed, dotted and dashed-dotted lines, respectively. }
	\label{fig-6}
\end{figure}

Fig.~\ref{fig-6} shows the temperature dependence of the ${ {}^{8}\text{Li} ( n , \gamma ) {}^{9}\text{Li} }$ reaction rate calculated using the neutron radiative capture cross-section from the GSM-CC calculation.
The total reaction rate, and the contribution of the capture to the ground, the first excited, and the first resonance are provided separately.
As seen in Fig.~\ref{fig-5}, the neutron capture cross section exhibits deviation from the usual $1/v$ behavior what has a direct impact on the temperature dependence of the reaction rate in Fig.~\ref{fig-6}.
The direct neutron capture to the ground state has the major contribution to the total reaction rate, in the range from $\sim82 $\% to $\sim92\%$.

The contribution of the first excited state $1/2^-_1$ was claimed to be negligible~\cite{Zecher1998,Kobayashi03,Garcia07}. We can see in Fig.~\ref{fig-6} and Table \ref{ratetab} that the capture to the first excited state $1/2^-_1$ could contribute to about $5 - 9\%$ of the total reaction rate, value which depends on temperature. The contribution of the $5/2^-_1$ resonance begins to increase significantly at $T_9 \approx 1$ and becomes more important than the contribution of $1/2^-_1$ state above $T_9 \approx 2$. At these temperatures, the resonance capture contribution can reach up to $10 \%$ of the total reaction rate.

%
%\begin{table}[ht!]
%\caption{The calculated reaction rate in units $cm^3 mol^{-1} s^{-1}$ for ${ {}^{8}\text{Li} ( n , \gamma ) {}^{9}\text{Li} }$ at $T_9=1$. The contributions of the capture cross section to the ground state $3/2^-_1$, the first excited state $1/2^-_1$ and the first resonant state $5/2^-_1$ in $^9$Li are given for a few values of the coefficient $f_{\rm E1}$.
% \label{ratetab}
% }
%\begin{ruledtabular}
%\begin{tabular}{p{1.5cm} p{1.5cm} cp{1.5cm}}
% $3/2^-_1$ & $1/2^-_1$& $5/2^-_1$\\
%    \hline
%  4619 &268  & 58  \\
%\end{tabular}
%\end{ruledtabular}
%\end{table}
%

\begin{table}[ht!]
\caption{The calculated reaction rate in units $cm^3 mol^{-1} s^{-1}$ for ${ {}^{8}\text{Li} ( n , \gamma ) {}^{9}\text{Li} }$ at $T_9=1$. The contributions of the capture cross section to the ground state $3/2^-_1$, the first excited state $1/2^-_1$ and the first resonant state $5/2^-_1$ in $^9$Li are given for a few values of the coefficient $f_{\rm E1}$.
 \label{ratetab}
 }
\begin{ruledtabular}
\begin{tabular}{p{0.6cm}|p{1.5cm}|p{1.5cm}|cp{1.5cm}}
  f$_{E1}$  &$3/2^-_1$ & $1/2^-_1$& $5/2^-_1$\\
    \hline
  1.0  &  4619 &268  & 58 \\
  1.2 & 5581& 377  &73  \\
  1.4 & 6778 &511 &99 \\
    1.6 &8159 & 666 &129 \\
     1.8 &9724 &841   &164  \\
\end{tabular}
\end{ruledtabular}
\end{table}

\begin{figure}[htbp]
	\includegraphics[width=1.00\linewidth]{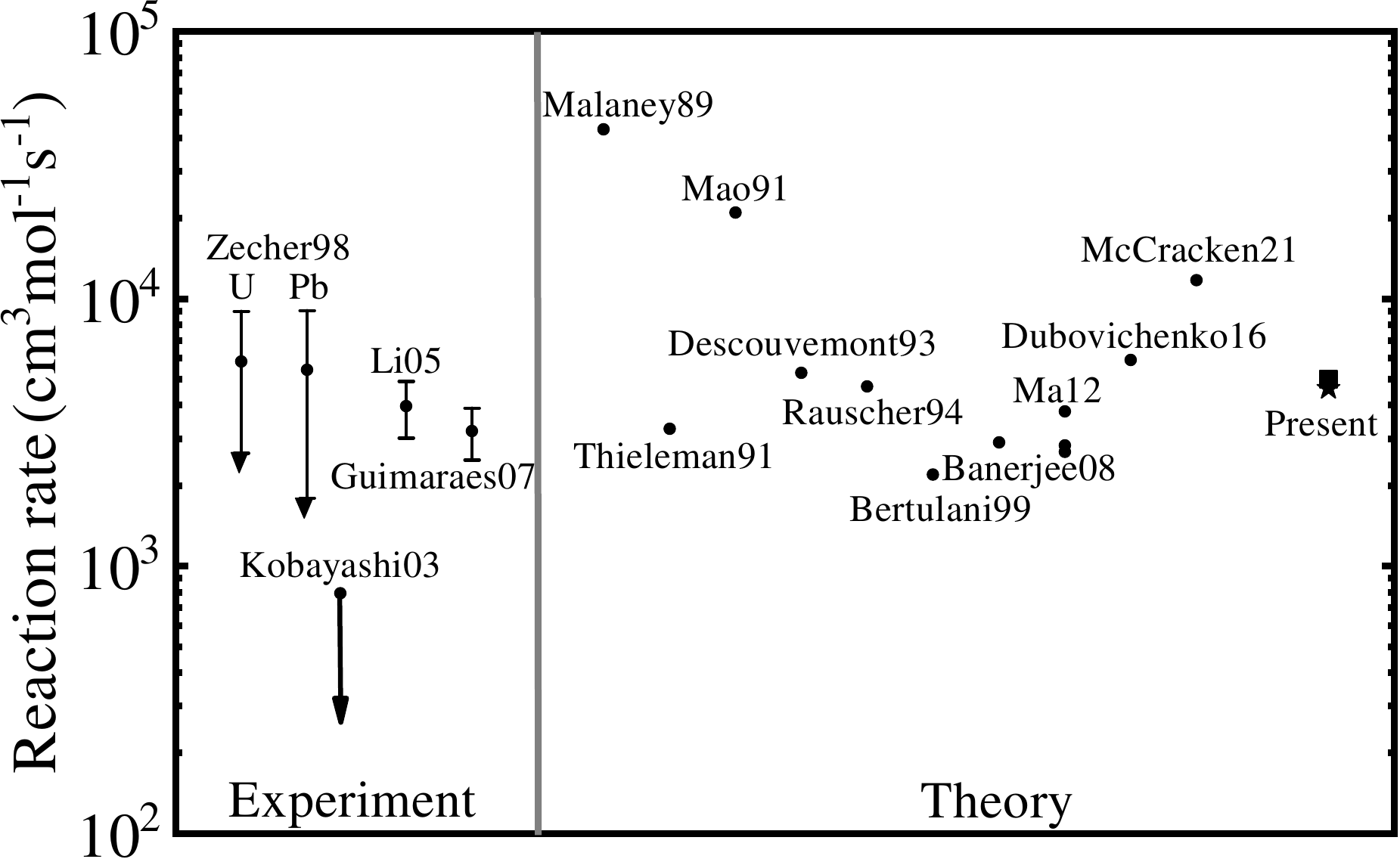}
	\caption{The comparison of experimental~\cite{Zecher1998,Kobayashi03,Garcia07,ZHLi05} and theoretical~\cite{Malaney89,Thielemann91,Mao91,Descouvemont93,Rauscher94,Bertulani99,Banerjee08,HLMa12,Dubovichenko16,NCSM21} reaction rates for the direct radiative neutron capture ${ {}^{8}\text{Li} ( n , \gamma ) {}^{9}\text{Li} }$ at $T_9=1$. The experimental upper limits in the Coulomb dissociation experiment with Pb and U targets~\cite{Zecher1998} are shown separately. The GSM-CC results are labeled as `Present'.
	The star and square symbols stand for the rate of the direct neutron capture to the ground state $3/2^-_1$ and total reaction rate obtained, respectively.	 }
	\label{fig-7}
\end{figure}
%

%The neutron capture reaction rate is shown in Table \ref{ratetab} for the temperature $T_9=1$ and separate contributions to the reaction rate from $3/2^-_1$, $1/2^-_1$ bound states and $5/2^-_1$ resonance.

The reaction rate depends on the value of the neutron effective charge. The dependence of the neutron capture reaction rate on the coefficient $f_{\rm E1}$ is shown  in Table \ref{ratetab} for the temperature $T_9=1$ and separate contributions to the reaction rate from $3/2^-_1$, $1/2^-_1$ bound states and $5/2^-_1$ resonance.
GSM-CC results satisfy the upper limit given in Ref.~\cite{Zecher1998} and is close to the rates: 3200$\pm$700 cm$^3$mol${^-1}$s$^{-1}$~\cite{Garcia07} and 3970$\pm$950 cm$^3$mol${^-1}$s$^{-1}$~\cite{ZHLi05}, which were obtained in the potential model using the SFs deduced from transfer reactions. %One may notice that these values can be reproduced by the GSM-CC with an E1 neutron effective charge whose empirical factor $f_{\rm E1}$ lies in the range $1.5 < f_{\rm E1} < 1.75$.

Various theoretical and experimental direct reaction rates for the reaction  ${ {}^{8}\text{Li} ( n , \gamma ) {}^{9}\text{Li} }$ at the temperature $T_9=1$ are shown in Fig.~\ref{fig-7}. For GSM-CC, we show the rates for the total neutron radiative capture cross section and its direct part. In the direct part of the neutron capture cross section, we neglect the contribution from the decay to first excited state $J^{\pi}=1/2^-_1$ which could not be measured experimentally~\cite{Zecher1998}. The star in Fig.~\ref{fig-7}
depicts the value of the direct neutron capture cross section for a standard value of E1 neutron effective charge
The square in Fig.~\ref{fig-7} represents the total neutron capture cross section, which includes contributions from the decays to $3/2^-_1$, $1/2^-_1$ bound states and $5/2^-_1$ resonance. One should stress that the GSM-CC prediction of the reaction rate shown in Fig.~\ref{fig-7} can be considered as the lower limit because the reaction rate increases with the scaling factor $f_{E1}$ (see Table \ref{ratetab}).

\section{Conclusions}
\label{sec4}

The ${ {}^{8}\text{Li} ( n , \gamma ) {}^{9}\text{Li} }$  reaction plays a pivotal role in the nucleosynthesis of elements heavier than carbon at $T_9 \approx 1$ in the inhomogeneous big-bang scenario. Heavier elements could also be produced at $0.5< T_9 <4.0$ using light neutron-rich nuclei in the high neutron abundance environment of the r-process in Type-II supernovae ~\cite{Gorres95,Terasawa01}. Also in this case, $^8$Li plays a key role in the subsequent synthesis of heavier elements.
As the direct experimental measurement of the rate for ${ {}^{8}\text{Li} ( n , \gamma ) {}^{9}\text{Li} }$ reaction is not possible, the information about considered astrophysical processes has to come from indirect methods, such as Coulomb dissociation or transfer reactions.
In fact, reliable theoretical estimates of this reaction rate are mandatory.

In this work  we applied GSM, which, in its coupled channels representation, provides a unified approach of nuclear structure and reactions. GSM-CC is a flexible microscopic model which can be applied to describe spectra and reaction chains in both heavy and light nuclei. Using a Hamiltonian with a two-body interaction adjusted to reproduce the spectra and binding energies of $^8$Li and $^{9}$Li, we calculated the neutron radiative capture cross section for the reaction ${ {}^{8}\text{Li} ( n , \gamma ) {}^{9}\text{Li} }$ and determined the temperature dependence of the reaction rate.

The GSM-CC calculation of $^9$Li provides a satisfactory description of all bound and resonance states with the excitation energy less than $\sim 6.5$ MeV. We are predicting the broad resonance $3/2^-_3$ which is overlapping with the known resonance $3/2^-_2$~\cite{Tilley04}. The energy difference of these two resonances is 110 keV and their widths have a comparable size (see Table \ref{spectra}). Hence, $3/2^-_2$ and $3/2^-_3$ resonances remind the avoided resonances in a vicinity of the exceptional point, where wave functions of these two different resonances coalesce~\cite{Zir83,Hei91,Hei91a,Brentano90,Brentano91,Okolowicz2011}.
At present, the properties of this doublet of resonances remain unknown since the energy and width of $3/2^-_3$ state have not yet been found experimentally. In this context, it is interesting to notice that the NCSMC calculation predicts the spacing of $\sim 1.2$ MeV for $3/2^-_2$ and $3/2^-_3$ resonances to be much larger than that found in GSM-CC. Moreover,  the width $\Gamma(3/2^-_3)$ in NCSMC is almost 3 times larger than the width found in GSM-CC.

The reaction rate obtained in the GSM-CC approach is consistent with the upper limit given by Zecher et al.~\cite{Zecher1998} and exceeds by a factor $\sim 5$ the limit reported by Kobayashi et al.~\cite{Kobayashi03}. Both these limits have been obtained using the Coulomb dissociation method combined with the detailed balance theorem. One order of magnitude  difference between the results of Coulomb dissociation experiments~\cite{Zecher1998,Kobayashi03} for the rate of direct neutron capture justifies the re-measurement of this reaction. Potential and cluster model calculations agree within $\sim30$\%. GSM-CC results for the reaction rate are close to the median value of results for potential models and cluster models.

%GSM-CC (NCSMC) results for the reaction rate are lower (higher) by $\sim 20$\% ($\sim30$\%) than the median value of results for potential models and cluster models.

%The major uncertainty of the GSM-CC calculation is related to the value of the used neutron effective charge. Using its theoretical value $e_{\rm eff}^n({\rm E1}) = eZ/A$, the GSM-CC reaction rate is lower than those extracted from transfer reactions~\cite{ZHLi05,Garcia07} and predicted by other theoretical approaches using various variants of the potential model~\cite{Mohr03,Mao91,HLMa12} or of the cluster model~\cite{Descouvemont93,Dubovichenko16}.

The GSM-CC rate of neutron capture reaction ${ {}^{8}\text{Li} ( n , \gamma ) {}^{9}\text{Li} }$ agrees with earlier estimates and indicates the destruction of $^8$Li in early universe. % is avoided.
Therefore, we also expect a reduction of the nucleosynthesis of heavier elements in the principal chain of reactions: $^8$Li($\alpha,n$)$^{11}$B($n,\gamma$)$^{12}$B($\beta^+$)$^{12}$C$\cdots$.

\section{Acknowledgements}
This work has been supported by the National Natural Science Foundation of China under Grant Nos. U2067205, 12175281, U1732138, and 12147219. G.X.D. and X.B.W. contributed equally to this work.

\end{document}